\documentclass[%
 reprint,
 amsmath,amssymb,
 aps,
 superscriptaddress,
]{revtex4-2}

\usepackage{graphicx}%
\usepackage{dcolumn}%
\usepackage{bm}%
\usepackage{hyperref}%

\newcommand{\MoSe}{\mathrm{MoSe_2}}
\newcommand{\FGT}{\mathrm{Fe_3GaTe_2}}

\begin{document}

\title{Magneto-optical Kerr spectroscopy of exciton Rydberg states in a magnetic van der Waals heterostructure}

\author{Astha Khandelwal}
\affiliation{Research Centre for Materials Nanoarchitectonics, National Institute for Materials Science, 1-1 Namiki, Tsukuba 305-0044, Japan}
\affiliation{Department of Materials Engineering, Indian Institute of Science, Bengaluru 560012, Karnataka, India}

\author{Benran Zhang}
\affiliation{Mork Family Department of Chemical Engineering and Materials Science, University of Southern California, Los Angeles, California 90089, USA}

\author{Madhusmita Jena}
\affiliation{Research Centre for Materials Nanoarchitectonics, National Institute for Materials Science, 1-1 Namiki, Tsukuba 305-0044, Japan}
\affiliation{Department of Physics, Indian Institute of Technology (B.H.U.), Varanasi 221005, India}

\author{Zhenchao Wen}
\affiliation{Research Centre for Magnetic and Spintronic Materials, National Institute for Materials Science, Tsukuba, Ibaraki, Japan}

\author{Kenji Watanabe}
\affiliation{Research Center for Electronic and Optical Materials, National Institute for Materials Science, 1-1 Namiki, Tsukuba 305-0044, Japan}

\author{Takashi Taniguchi}
\affiliation{Research Centre for Materials Nanoarchitectonics, National Institute for Materials Science, 1-1 Namiki, Tsukuba 305-0044, Japan}

\author{Saroj P. Dash}
\affiliation{Department of Microtechnology and Nanoscience, Chalmers University of Technology, Gothenburg, SE-41296, Sweden}

\author{Zhenglu Li}
\affiliation{Mork Family Department of Chemical Engineering and Materials Science, University of Southern California, Los Angeles, California 90089, USA}

\author{Ryo Kitaura}
\affiliation{Research Centre for Materials Nanoarchitectonics, National Institute for Materials Science, 1-1 Namiki, Tsukuba 305-0044, Japan}
\affiliation{Graduate School of Chemical Sciences and Engineering, Hokkaido University, Sapporo, 060-8628, Japan}

\author{Bhagwati Prasad}
\email{bpjoshi@iisc.ac.in}
\affiliation{Department of Materials Engineering, Indian Institute of Science, Bengaluru 560012, Karnataka, India}

\author{Daichi Kozawa}
\email{KOZAWA.Daichi@nims.go.jp}
\affiliation{Research Centre for Materials Nanoarchitectonics, National Institute for Materials Science, 1-1 Namiki, Tsukuba 305-0044, Japan}
\affiliation{Department of Materials Science, Institute of Pure and Applied Sciences, University of Tsukuba, Tsukuba 305-8573, Japan}

\begin{abstract}
Excitonic states in two-dimensional semiconductors are sensitive to time-reversal symmetry breaking, yet how interfacial exchange acts on higher-lying exciton Rydberg states remains largely unexplored. Here we show that wavelength-resolved magneto-optical Kerr spectroscopy of a proximity-coupled $\MoSe /\FGT$ van der Waals heterostructure resolves magnetic symmetry breaking across the A- and B-exciton manifolds. Combining reflection, photoluminescence, and photoluminescence excitation spectroscopy with first-principles many-body calculations including screening by the metallic $\FGT$ layer, we assign the strongly enhanced Kerr resonances to the ground and 2s states of both manifolds. Reversing the magnetic field inverts the polarity of every resonance. A field-odd/field-even decomposition confirms a magnetic origin. The opposite Kerr polarities of the A- and B-exciton manifolds can be explained by proximity-induced exchange coupling to the opposite valence-band spins of the two exciton series. Our results establish resonant Kerr spectroscopy as a sensitive probe of magnetically induced symmetry breaking across the excitonic manifold.
\end{abstract}

\maketitle

\section{Introduction}

Breaking inversion and time-reversal symmetries underpins the emergence of topological and nonreciprocal phenomena across a wide range of quantum materials, from oxide superlattices and thin-film heterostructures to van der Waals stacks~\cite{Nadeem2023,Lee2005,Hwang2012}. Engineered asymmetry in such systems gives rise to direction-dependent transport and optical responses, including nonlinear Hall effects~\cite{Sodemann2015}, superconducting diode behavior~\cite{Ando2020,Davydova2024}, and magneto-chiral anisotropy~\cite{Du2024,Hou2023}. Among these platforms, van der Waals heterostructures offer a uniquely controllable route to breaking time-reversal symmetry without external magnetic fields: magnetic proximity coupling induces exchange interactions in an adjacent nonmagnetic layer across an atomically sharp interface, lifting its spin degeneracy, as observed in graphene~\cite{Hu2024,Zheng2022,Karpiak_2019}.

Proximity exchange has a particularly strong effect on monolayer transition metal dichalcogenides (TMDCs), whose optical response is dominated by tightly bound excitons~\cite{Cao2012,Shahnazaryan2017,Dirnberger2022}. An adjacent magnetic layer lifts the valley degeneracy of the band edges~\cite{Bora2021,Huang2020}, observed as pronounced valley splitting and spin polarization in excitonic photoluminescence~\cite{Wilson2021,Lan2026,Li_2023}. In $\MoSe/\mathrm{CrBr_3}$ heterostructures, spin-dependent hybridization further shifts the $K$ and $K'$ excitons asymmetrically, beyond a simple effective magnetic-field description~\cite{Choi_2022}. Beyond photoluminescence, magneto-optical Kerr effect (MOKE) spectroscopy is a sensitive probe of the resulting broken time-reversal symmetry, because Kerr rotation arises from a finite off-diagonal optical conductivity enabled by the broken symmetry~\cite{Ogawa2003,Trepanier2022,Fried2014}; MOKE is therefore widely used to study magnetic order, hysteresis, and domain formation in atomically thin magnets and their heterostructures~\cite{Huang2017,Jiang2018,Gong2017}. Yet most MOKE studies of TMDC/two-dimensional magnet heterostructures rely on a single wavelength, leaving the role of excitonic resonances in the Kerr response unclear, even though the exciton contribution is expected to be pronounced owing to the interplay between exchange interactions and many-body Coulomb effects~\cite{Henriques2020,Wang2018,Mak2014,Zollner2019}. Resonant magnetic circular dichroism and magneto-reflectance have probed the excitonic transitions in related heterostructures~\cite{Norden2019,Seyler2018}, but how proximity-induced exchange manifests across the higher-lying states and the full A- and B-exciton manifolds remains an open question.

Here we investigate the excitonic magneto-optical Kerr response of a proximity-coupled, hBN-encapsulated $\MoSe/\FGT$ van der Waals heterostructure. We acquire wavelength-resolved Kerr rotation spectra across the A- and B-exciton manifolds and assign the spectral features to neutral excitons and their higher-lying states by combining reflection, photoluminescence (PL), and photoluminescence excitation (PLE) spectroscopy. To separate the magnetic response from non-magnetic contributions, we measure the Kerr spectra at opposite magnetic fields and decompose them into field-odd and field-even components, and we examine the Kerr polarities of the A- and B-exciton manifolds in terms of proximity-induced exchange. Finally, we compare the assignment with first-principles GW plus Bethe-Salpeter equation (GW-BSE) calculations in which the dielectric screening of the metallic $\FGT$ layer is varied systematically.

\section{Results}

\begin{figure}
\includegraphics[width=\columnwidth]{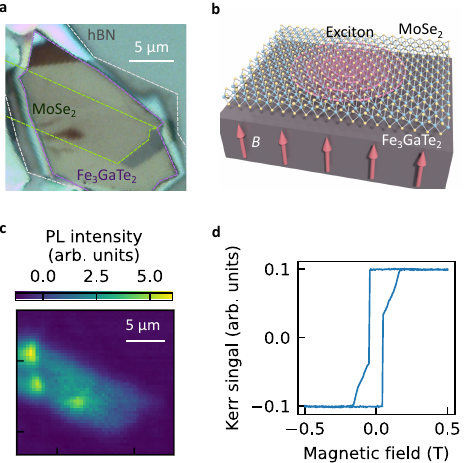}
\caption{\label{fig:fig1}Device structure and magnetic characterization of the $\MoSe/\FGT$ heterostructure. \textbf{a}, Optical microscope image of the hBN-encapsulated heterostructure, highlighting the spatial overlap between the monolayer $\MoSe$ and the multilayer $\FGT$ flakes. \textbf{b}, Schematic representation of magnetic proximity coupling at the interface. The out-of-plane magnetization of $\FGT$ induces an exchange interaction in the $\MoSe$ monolayer, which can lift the degeneracy of excitonic transitions at the $K$ and $K'$ valleys~\cite{Lyons2020,Beer2024}. \textbf{c}, PL intensity map used to identify the monolayer $\MoSe$ region. The PL signal delineates the monolayer area, while partial quenching is observed in the region overlapping with the metallic $\FGT$. \textbf{d}, Kerr signal of the $\FGT$ flake as a function of the applied magnetic field, exhibiting a hysteresis loop. All measurements shown in Fig.~\ref{fig:fig1} are conducted at room temperature.}
\end{figure}

We prepare hBN-encapsulated $\MoSe/\FGT$ heterostructures by mechanical exfoliation followed by deterministic transfer~\cite{Liu2019,Geim2013} (see Methods), then characterize their configuration together with the magnetism of the $\FGT$ layer (Fig.~\ref{fig:fig1}). The multilayer hBN encapsulation protects the heterostructure from environmental degradation and ensures a clean interface. Figure~\ref{fig:fig1}a shows an optical microscope image of the assembled heterostructure, highlighting the spatial overlap between the $\MoSe$ monolayer and the magnetic flake. As illustrated schematically in Fig.~\ref{fig:fig1}b, the $\FGT$ layer exhibits out-of-plane magnetization and can induce an exchange interaction in the $\MoSe$ monolayer through magnetic proximity coupling, thereby linking the excitonic optical response of $\MoSe$ to the interfacial magnetism~\cite{Wu2019}. The monolayer region is identified using the PL intensity map shown in Fig.~\ref{fig:fig1}c, in which the PL signal clearly delineates the $\MoSe$ monolayer, although partial quenching is observed in the region overlapping with metallic $\FGT$, likely because of enhanced nonradiative charge- and/or energy-transfer processes~\cite{Lorchat2020,Froehlicher2018}. The monolayer character of $\MoSe$ is independently confirmed by Raman spectroscopy (Supplementary Fig.~S1). The magnetic properties of the $\FGT$ flake are characterized using MOKE measurements. The Kerr signal as a function of the applied magnetic field (Fig.~\ref{fig:fig1}d) exhibits a clear hysteresis loop that confirms the ferromagnetic ordering of $\FGT$.

\begin{figure*}
\includegraphics[width=0.85\textwidth]{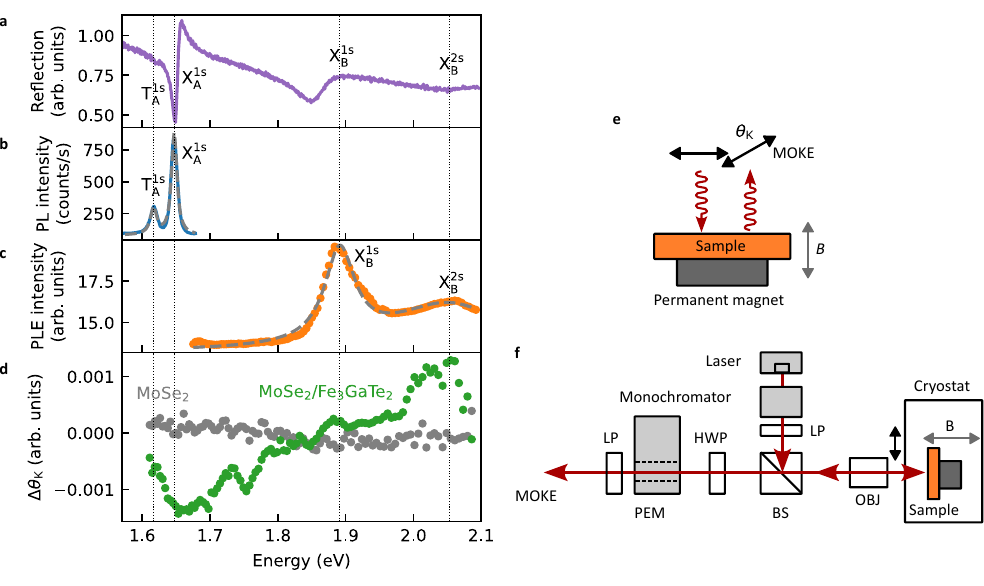}
\caption{\label{fig:fig2}Excitonic resonances and magneto-optical Kerr response in the $\MoSe/\FGT$ heterostructure. Labels mark the excitonic resonances, and vertical dotted lines across panels \textbf{a}--\textbf{d} indicate the resonance energies extracted from the PL and PLE spectra. \textbf{a}, Reflection spectrum of the $\MoSe/\FGT$ heterostructure. \textbf{b}, PL spectrum of the $\MoSe/\FGT$ heterostructure under excitation at 1.96~eV (blue curve). \textbf{c}, PLE spectrum of the heterostructure (orange dots). In \textbf{b} and \textbf{c}, gray broken curves show multi-Lorentzian fits using Eq.~(\ref{eq:lorentzian}). \textbf{d}, Differential Kerr rotation spectrum ($\Delta\theta_\mathrm{K}$) under an out-of-plane magnetic field of $+300$~mT. Gray dots show the control differential MOKE spectrum of monolayer $\MoSe$ on a separate device, referenced to the bare Si substrate; green dots show the differential spectrum of the $\MoSe/\FGT$ heterostructure, referenced to a bare $\FGT$ region [Eq.~(\ref{eq:diffkerr})]. \textbf{e}, Schematic of the magneto-optical Kerr effect: linearly polarized incident light reflected from the out-of-plane magnetized sample acquires a rotation of its polarization plane by the Kerr angle $\theta_\mathrm{K}$. \textbf{f}, Schematic of the experimental setup for MOKE spectroscopy. LP, linear polarizer; HWP, half-wave plate; PEM, photoelastic modulator; BS, beam splitter; OBJ, objective lens. The PEM is electrically coupled to a lock-in amplifier. All the measurements shown in Fig.~\ref{fig:fig2} are conducted at 40~K with a laser power below 50~$\mu$W.}
\end{figure*}

The excitonic resonances of the $\MoSe/\FGT$ van der Waals heterostructure are next identified using reflection spectroscopy (Fig.~\ref{fig:fig2}a) and PL spectroscopy (Fig.~\ref{fig:fig2}b). The reflection spectrum shows the negatively charged trion $T_A^{1s}$ and the neutral exciton $X_A^{1s}$ together with the B-exciton ground state $X_B^{1s}$ and 2s state $X_B^{2s}$, although the higher-lying exciton states appear only as weak features. The PL spectrum exhibits two pronounced emission peaks centered at approximately 1.617~eV and 1.647~eV, corresponding to $T_A^{1s}$ and $X_A^{1s}$, respectively. The prominent spectral weights of these resonances demonstrate that the PL response is dominated by excitonic transitions in monolayer $\MoSe$. To quantitatively analyze the spectral features, the PL spectra are fitted using a multi-Lorentzian function with a constant background,
\begin{equation}
\sum_{i=1}^{N}\frac{a_i\,(w_i/2)^{2}}{(E-E_i)^{2}+(w_i/2)^{2}}+C,
\label{eq:lorentzian}
\end{equation}
where $E$ is the photon energy, $N$ is the number of excitonic resonances included in the fit, and $i$ labels the individual optical transitions. The parameters $a_i$, $E_i$, and $w_i$ denote the amplitude, resonance energy, and full width at half maximum of the $i$-th resonance, respectively, while $C$ represents a constant offset background. The PL spectrum is fitted with Eq.~(\ref{eq:lorentzian}) using $N=2$, corresponding to the neutral exciton $X_A^{1s}$ and negatively charged trion $T_A^{1s}$. The extracted linewidths are approximately 9.6~meV for $X_A^{1s}$ and 11~meV for $T_A^{1s}$, consistent with high optical quality of the heterostructure.

PLE spectroscopy further resolves higher-energy resonances associated with the B-exciton manifold (Fig.~\ref{fig:fig2}c), observed near 1.891~eV and 2.053~eV and assigned to $X_B^{1s}$ and $X_B^{2s}$, respectively, consistent with the expected excitonic hierarchy and spin-orbit-split valence bands~\cite{Wang2015,Han2018}. The fits yield resonance energies and linewidths that form the basis for interpreting the magneto-optical Kerr response in the heterostructure.

We then measure the magneto-optical response of the $\MoSe/\FGT$ heterostructure with a custom-built MOKE spectroscopy setup at 40~K (see Methods). The measurement temperature of 40~K is chosen so that the neutral exciton and negatively charged trion remain spectrally resolved, as shown by temperature-dependent PL (Supplementary Fig.~S2). In MOKE, linearly polarized light reflected from a magnetized sample undergoes a rotation in its polarization plane by the Kerr angle $\theta_\mathrm{K}$, as sketched in Fig.~\ref{fig:fig2}e. Under the present measurement configuration (Fig.~\ref{fig:fig2}f), the Kerr rotation is proportional to the AC signal detected with a lock-in amplifier, normalized by the DC reflected intensity. We tune the excitation wavelength across the optical resonances under an out-of-plane magnetic field of $+300$~mT, which is sufficient to saturate the $\FGT$ magnetization. We collect MOKE spectra at three distinct locations and calculate their average. To isolate the excitonic contribution of the $\MoSe/\FGT$ heterostructure, we subtract the Kerr spectrum of a bare $\FGT$ region from that of the heterostructure; as a control, we subtract the Si-substrate spectrum from that of monolayer $\MoSe$ measured on a separate device. The differential Kerr response of the heterostructure is thus
\begin{equation}
\Delta\theta_\mathrm{K}(E)=\theta_\mathrm{K}^{\MoSe/\FGT}(E)-\theta_\mathrm{K}^{\FGT}(E),
\label{eq:diffkerr}
\end{equation}
where $\theta_\mathrm{K}^{\MoSe/\FGT}(E)$ and $\theta_\mathrm{K}^{\FGT}(E)$ denote the energy-dependent Kerr rotations measured on the heterostructure and reference regions, respectively.

The averaged differential spectrum of the $\MoSe/\FGT$ heterostructure shows pronounced resonant Kerr features at the excitonic transitions identified by reflection, PL, and PLE (Fig.~\ref{fig:fig2}d), demonstrating the strong excitonic contribution to the magneto-optical response in monolayer $\MoSe$. Because Kerr rotation originates from broken time-reversal symmetry through a finite off-diagonal optical conductivity, the differential signal directly reflects proximity-induced exchange interactions in the semiconductor layer. The Kerr spectra show resonances associated with $X_A^{1s}$, the higher-lying A-exciton state, and the B-exciton manifold, with the A- and B-exciton series exhibiting opposite Kerr polarities that indicate distinct spin-valley character of the underlying optical transitions. The excited 2s excitonic states are resolved more clearly in the Kerr response than in the reflection, PL, and PLE spectra, in which they appear only as weak features.

To quantify the Kerr resonances, we fit $\Delta\theta_\mathrm{K}$ with Eq.~(\ref{eq:lorentzian}) phenomenologically using $N=4$, corresponding to the resonances $X_A^{1s}$, $X_A^{2s}$, $X_B^{1s}$, and $X_B^{2s}$. The number of components is selected by the Akaike and Bayesian information criteria, both of which favor four over larger sets. Only the resonance energies and linewidths are bounded; the amplitudes, including their signs, are free parameters, so the polarity of each resonance is an outcome of the fit rather than a constraint imposed on it.

The extracted resonance energies reveal a structured excitonic hierarchy in the Kerr response. The A-exciton series $X_A^{1s}$, $X_A^{2s}$ follows a Rydberg-like progression characteristic of reduced dielectric screening and enhanced Coulomb interactions in two-dimensional semiconductors~\cite{Liu_2021,Chernikov2014}. The B-exciton resonances $X_B^{1s}$, $X_B^{2s}$ appear at higher energies owing to spin-orbit splitting of the valence band~\cite{Xiao2012}. We do not resolve a trion feature in the Kerr spectra: a trion component with free position, amplitude and width is not determined at both field polarities, and the excitation range below 1.61~eV is too sparsely sampled to constrain one. The $X_A^{1s}$ resonance in the Kerr spectra also appears at a slightly lower energy than in PL. This small difference may arise from the absorption-like line shape of the Kerr resonances, in contrast to the emission line shape measured in PL, as well as from local variations in the dielectric environment between the sampled positions, rather than from a conventional Stokes shift.

\begin{figure*}
\includegraphics[width=0.8\textwidth]{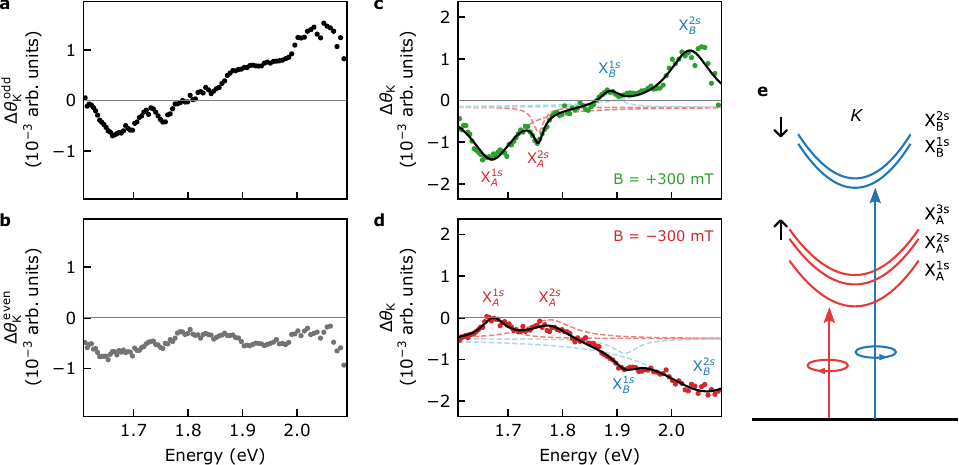}
\caption{\label{fig:fig3}Field-symmetry decomposition and multi-Lorentzian analysis of the differential Kerr rotation spectra, measured at 40~K with a laser power below 50~$\mu$W. Throughout, $\Delta\theta_\mathrm{K}$ denotes the Kerr rotation of the $\MoSe/\FGT$ heterostructure referenced to a bare $\FGT$ region [Eq.~(\ref{eq:diffkerr})]. \textbf{a}, Field-odd component, $\Delta\theta_\mathrm{K}^{\,\mathrm{odd}} = [\Delta\theta_\mathrm{K}(+B) - \Delta\theta_\mathrm{K}(-B)]/2$. \textbf{b}, Field-even component, $\Delta\theta_\mathrm{K}^{\,\mathrm{even}} = [\Delta\theta_\mathrm{K}(+B) + \Delta\theta_\mathrm{K}(-B)]/2$, plotted on the same vertical scale as \textbf{a}. It consists of a constant offset with no resonant structure. \textbf{c},\textbf{d}, Differential Kerr spectra under out-of-plane magnetic fields of $+300$~mT (\textbf{c}, green) and $-300$~mT (\textbf{d}, red), on matched vertical scales. Dashed curves denote the individual Lorentzian components and the solid black curve the total fit using Eq.~(\ref{eq:lorentzian}), with the amplitude signs as free parameters. The model comprises four neutral-exciton components plus a constant offset, with no trion and no 3s component: $X_A^{1s}$, $X_A^{2s}$, $X_B^{1s}$ and $X_B^{2s}$, each labeled at the energy fitted for that field direction. \textbf{e}, Schematic energy-level diagram of the A- and B-exciton Rydberg series at the $K$ point. The red arrow denotes the optical transition to the A-exciton and the blue arrow the transition to the B-exciton; the two derive from valence sub-bands of opposite spin, as indicated by the black arrows. The $3s$ level is drawn for completeness of the series and is not resolved in these spectra.}
\end{figure*}

Measurements at opposite magnetic fields establish that the Kerr resonances are magnetic in origin (Fig.~\ref{fig:fig3}). To separate the magnetic response from any contribution that does not reverse with the magnetization, we decompose the spectra measured at the two field directions into field-odd and field-even components, $\Delta\theta_\mathrm{K}^{\,\mathrm{odd}} = [\Delta\theta_\mathrm{K}(+B) - \Delta\theta_\mathrm{K}(-B)]/2$ and $\Delta\theta_\mathrm{K}^{\,\mathrm{even}} = [\Delta\theta_\mathrm{K}(+B) + \Delta\theta_\mathrm{K}(-B)]/2$ (Figs.~\ref{fig:fig3}a and~\ref{fig:fig3}b). The field-odd component carries the excitonic structure in full, whereas the field-even component consists of a constant offset with no resonant feature. This establishes, without recourse to any fitting, that the resonances originate from the magnetization rather than from a static instrumental or reflectivity contrast. Figures~\ref{fig:fig3}c and~\ref{fig:fig3}d show the corresponding differential Kerr rotation spectra measured under out-of-plane magnetic fields of $+300$~mT and $-300$~mT, respectively. The spectra exhibit pronounced resonant features associated with the A- and B-exciton manifolds. The curvature of each Kerr resonance changes sign when the field is reversed, consistent with the field-odd decomposition. Resonance energies, obtained from multi-Lorentzian fits [Eq.~(\ref{eq:lorentzian})] that separate the individual optical resonances across the full spectral range, are quoted from the $+300$~mT spectrum, in which the excitonic features are most clearly resolved against the non-resonant background; the fitted positions in the $-300$~mT spectrum are correspondingly less well constrained (see Supplementary Note~\mbox{I-A}).

Although the fitting reliably captures the spectral positions and relative polarities of the Kerr resonances, the absolute amplitudes should be interpreted with caution. The measured Kerr signal is influenced not only by the intrinsic optical response of the $\MoSe$ layer, but also by additional optical and structural factors, including spectral overlap with the Kerr response of the underlying $\FGT$, local thickness variations of the ferromagnetic layer, optical interference within the multilayer heterostructure~\cite{Sumi2018}, reflectivity contrast between different regions, and absorption associated with the metallic nature of $\FGT$. These contributions are not a secondary correction: the ratio of the fitted amplitudes at the two field polarities is not common to the four resonances, whereas a uniform rescaling of the magnetization, of the detector response or of the optical gain would produce a single ratio for every one of them. The spread indicates an energy-dependent optical contribution, which is expected for a Kerr measurement on a multilayer and is quantified in Supplementary Note~\mbox{I-B}. The analysis therefore relies on the resonance energies and the sign structure of the Kerr features, which are unaffected by a positive, slowly varying optical factor and provide the most robust signatures of proximity-induced modifications to the optical transitions in the heterostructure.

The A- and B-exciton resonances display opposite curvature (Figs.~\ref{fig:fig3}c and~\ref{fig:fig3}d), a polarity that follows from the spin structure of the transitions and is a signature of exchange coupling rather than of the applied field. At the $K$ point the valence band is split by spin-orbit coupling into two sub-bands of opposite spin, and the A and B excitons are the spin-conserving transitions built on the upper and lower sub-band, respectively~\cite{Xiao2012,Xu2014}. An external magnetic field cannot distinguish them: for a spin-conserving transition the spin Zeeman shifts of the electron and the hole cancel in the transition energy, leaving orbital and valley contributions that are the same for both manifolds. Magneto-reflectivity measurements on monolayer $\MoSe$ accordingly find valley Zeeman splittings that are equal for the A and B excitons within experimental accuracy, with $g \approx -4$~\cite{Koperski_2018}; at $300$~mT this corresponds to a splitting of only $0.07$~meV. Proximity exchange instead couples to spin, so its contributions do not cancel. The induced valley splitting is set by the difference between the exchange splittings of the conduction and valence bands and reverses sign between the A and B manifolds, because they are built on valence states of opposite spin. An excitonic theory of proximity-induced Kerr rotation in monolayer TMDCs reproduces this explicitly, giving valley splittings of equal magnitude and opposite sign for the two spin-orbit-split manifolds~\cite{Henriques2020} (see Supplementary Note~\mbox{I-C}). The opposite polarity observed here is therefore evidence that the Kerr response is governed by interfacial exchange rather than by the applied field, which alone would shift both manifolds in the same direction. The sign structure is preserved across the higher-lying states~\cite{Henriques2020}, indicating that the Kerr response retains the underlying spin-valley character of the electronic band structure.

\begin{figure*}
\includegraphics[width=0.6\textwidth]{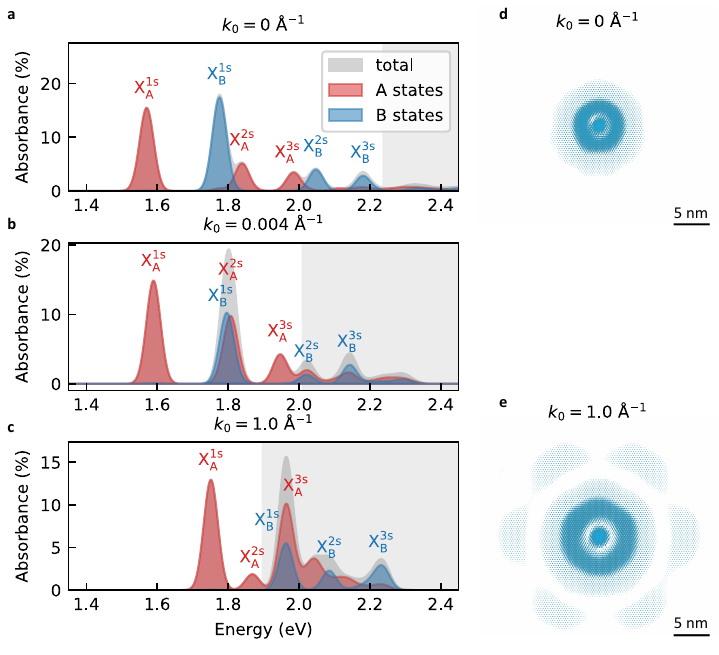}
\caption{\label{fig:fig4}Screening-dependent excitonic absorption of monolayer $\MoSe$. \textbf{a}--\textbf{c}, Optical absorption spectra calculated by GW-BSE with dielectric screening mimicking the effects from the metallic $\FGT$ layer, for screening strengths $k_0=0$ (\textbf{a}), $0.004$ (\textbf{b}) and $1.0\,\text{\AA}^{-1}$ (\textbf{c}). Gray filled curves show the total absorbance, and red and blue curves the contributions of A- and B-character states, assigned from Bloch band projections. The light-gray shaded region marks the electron-hole continuum above the quasiparticle gap; peaks inside it are labeled by the bound state from which they evolve (e.g.\ $X_A^{2s}$-derived). With increasing $k_0$ the higher-lying states shift and reorder, including an $X_A^{2s}$/$X_B^{1s}$ crossing, and broaden as they approach the continuum. \textbf{d},\textbf{e}, Real-space distribution of the electron part of the exciton wavefunction, with the hole fixed at the center, for the $X_A^{2s}$ state at $k_0=0$ (\textbf{d}) and the $X_A^{2s}$-derived state at $k_0=1.0\,\text{\AA}^{-1}$ (\textbf{e}). Scale bars, 5~nm.}
\end{figure*}

To support the assignment of the higher-lying excitonic resonances observed in the MOKE spectra, we perform first-principles calculations of the optical absorption spectrum within the GW-BSE framework while incorporating dielectric screening mimicking the effects from the metallic $\FGT$ layer. The screening effect of $\FGT$ is modeled as a freestanding two-dimensional electron gas in the static limit and implemented through a wavevector-dependent correction to the dielectric matrix. The screening strength is parameterized by the inverse screening length $k_0$, with larger $k_0$ corresponding to stronger dielectric screening. Varying $k_0$ systematically, we observe a reordering of the exciton peaks. As shown in Fig.~\ref{fig:fig4}a, in the absence of $\FGT$ screening at $k_0 = 0$, the excitonic resonances follow the conventional ordering expected for monolayer $\MoSe$. As $k_0$ increases (Fig.~\ref{fig:fig4}b,c), the $X_A^{2s}$ state crosses below the $X_B^{1s}$ state, in agreement with experiment. For $k_0 \approx 1\,\text{\AA}^{-1}$, the calculated ordering of the excitonic resonances qualitatively reproduces the sequence observed in the MOKE spectra (Fig.~\ref{fig:fig2}). This agreement provides strong support for our assignment of the higher-energy spectral features to higher-lying states of the A- and B-exciton Rydberg series and indicates that the dielectric screening from the underlying $\FGT$ layer plays a dominant role in reconstructing the excitonic spectrum of monolayer $\MoSe$. Figures~\ref{fig:fig4}d and~\ref{fig:fig4}e show the real-space exciton wavefunctions, which demonstrate a screening-induced delocalization of the exciton envelopes. Here, the radius of the wavefunction of the $X_A^{2s}$ state is $\sim$3.5~nm in the freestanding calculation and $\sim$6.4~nm in the calculation at $k_0 = 1\,\text{\AA}^{-1}$. The full screening-dependent absorption map, showing the A- and B-series excitons across the range of $k_0$, is provided in Supplementary Fig.~S3; the spectra in Fig.~\ref{fig:fig4} correspond to representative cases within this map.

Dielectric screening by the $\FGT$ layer determines not only the energetic ordering but also the spatial character of the excitonic states in $\MoSe/\FGT$ heterostructures. The screening-induced delocalization reflects the reduced Coulomb attraction between the composite electron and hole within an exciton, and is most pronounced for the higher-lying excitonic states. Because dielectric screening also reduces the quasiparticle band gap, for sufficiently strong screening ($k_0 > 0.3\,\text{\AA}^{-1}$) the $X_A^{2s}$ and higher states lie above the renormalized gap and are no longer truly bound excitons. We therefore label these features as $X_A^{2s}$-derived (and higher A-exciton-derived) states rather than as strict 2s or 3s excitons. Proximity exchange, in contrast, acts on the constituent electron and hole bands rather than through the exciton envelope. All states of a Rydberg series therefore inherit, at leading order, the same exchange-induced valley splitting, which is consistent with the higher-lying exciton features resolved in the Kerr response.

The present Kerr measurement complements magnetic circular dichroism and resolves the higher-lying 2s states of both exciton manifolds in the magneto-optical response. Circular dichroism probes the imaginary part of the magneto-optical conductivity, whereas the Kerr rotation is governed by its real part; excitonic effects qualitatively alter the line shapes of both relative to the single-particle picture~\cite{Henriques2020}. The Kerr resonances are accordingly absorption-like rather than dispersive, consistent with the multi-Lorentzian description of Eq.~(\ref{eq:lorentzian}). Such 2s resolution has not been achieved in previous circular-dichroism studies of proximity-coupled heterostructures.

\section{Conclusion}

In conclusion, we have investigated excitonic resonances in the magneto-optical Kerr response of a proximity-coupled $\MoSe/\FGT$ van der Waals heterostructure. Optical microscopy, PL mapping, and Kerr hysteresis measurements confirm the formation of a high-quality heterostructure with robust ferromagnetic ordering in the underlying $\FGT$ layer. Through reflection, PL, and PLE spectroscopy, we resolve the neutral exciton $X_A^{1s}$, trion $T_A^{1s}$, and higher-lying excitonic resonances associated with the A- and B-exciton manifolds. Wavelength-resolved Kerr spectroscopy reveals pronounced resonant magneto-optical features whose spectral positions coincide with the excitonic transitions identified by reflection, PL, and PLE, demonstrating the excitonic character of the Kerr response. The A- and B-exciton manifolds exhibit opposite Kerr polarities that reverse with the magnetic field; a field-odd/field-even decomposition confirms the magnetic origin of the resonances, and the opposite polarities are consistent with proximity-induced exchange coupling to the opposite valence-band spins of the two exciton series. First-principles GW-BSE calculations that include dielectric screening mimicking the effects from the metallic $\FGT$ layer show that increasing screening reorders the higher-lying excitons, with the $X_A^{2s}$ state crossing below $X_B^{1s}$, and reproduce the experimental ordering at a screening strength $k_0 \approx 1~\mathrm{\mathring{A}}^{-1}$, supporting the assignment of the Kerr resonances to the 1s and 2s states of both manifolds. The excited 2s states of both manifolds are resolved in the Kerr response, indicating that proximity-induced exchange interactions extend to the higher-lying excitonic states. These results show that resonant excitonic Kerr spectroscopy is a sensitive probe of proximity-induced magnetic interactions and spin-valley-dependent magneto-optical phenomena in van der Waals heterostructures, providing a pathway toward tunable optical and spintronic functionalities in two-dimensional quantum materials.

\section{Methods}

\subsection{Sample preparation}

$\FGT$ single crystals are obtained from 2D Semiconductors. Monolayer $\MoSe$ and hBN with a thickness range of 8-10~nm are obtained by mechanical exfoliation from bulk single crystals onto Si substrates. The monolayer nature of $\MoSe$ is confirmed by optical contrast, PL mapping, and Raman spectroscopy (see Supplementary Fig.~S1). Few-layer $\FGT$ flakes are mechanically exfoliated onto Si substrates inside a glove box with oxygen and moisture levels below 0.1~ppm, because $\FGT$ degrades rapidly under ambient conditions. Suitable $\FGT$ flakes are identified based on optical contrast, favoring regions with uniform thickness, smooth surfaces, and the absence of cracks or visible defects to ensure good contact with $\MoSe$. The magnetic properties of the $\FGT$ flakes are characterized by MOKE hysteresis loop measurements.

The hBN/$\MoSe$/$\FGT$ heterostructures are assembled using a dry pickup and transfer technique with an Elvacite polymer-coated glass stamp~\cite{Onodera2022}. The hBN/$\MoSe$ stack is first picked up onto the polymer stamp and subsequently aligned with the target $\FGT$ flake. The stack is brought into contact with the $\FGT$ at 120~$^{\circ}$C and held for 10-12~min to promote adhesion. After cooling to room temperature, the heterostructure remains attached to the substrate. The polymer is then removed by immersing the sample in chloroform for 12~h, yielding a clean hBN/$\MoSe$/$\FGT$ heterostructure on the Si substrate.

\subsection{Optical spectroscopy and imaging}

A home-built confocal microscope is used to collect reflection and PL spectra and to perform PL excitation imaging. The sample is mounted in a closed-cycle helium cryostat, and an objective lens with a numerical aperture of 0.42 is positioned using a motorized three-dimensional translation stage. The excitation polarization is set with a Glan-Taylor polarizer and a half-wave plate, with the incident polarization kept parallel to the optical table. The focal position is optimized by scanning the objective along the optical axis to maximize the reflected signal. Reflection spectra are acquired using a supercontinuum white-light laser, whereas PL spectroscopy and imaging use the same laser coupled to a 15-cm monochromator that selects a narrow excitation band. The excitation linewidth is approximately 5~nm, which sets the spectral resolution of the excitation energy in both the PL excitation spectroscopy and the MOKE spectroscopy. For PL excitation spectroscopy, the wavelength is tuned with the monochromator and the excitation beam is spectrally filtered with a short-pass filter before being focused onto the sample. The signal is collected through the same objective in a backscattering geometry, passed through a long-pass filter, and analyzed using a 328-mm spectrometer equipped with a thermoelectrically cooled silicon photodiode array detector. The excitation and detection wavelengths are calibrated against Hg and Ne-Ar spectral lamps prior to the measurements.

Excitation imaging is performed to characterize the spatial distribution of the $\MoSe$ PL intensity at room temperature. An objective lens with a numerical aperture of 0.65 is used, and the PL intensity is collected on a Si avalanche photodiode. A raster scan is performed over the area containing the $\MoSe$, recording the emission intensity at each position.

\subsection{Magneto-optical Kerr effect spectroscopy}

Wavelength-dependent MOKE measurements are performed using the same cryostat and optical setup employed for the PLE experiments (Fig.~\ref{fig:fig2}f). A supercontinuum laser coupled to a 15-cm monochromator serves as the tunable excitation source. The excitation beam is linearly polarized with a Glan-Thompson calcite polarizer. The polarization angle is adjusted using a half-wave plate before the beam passes through a photoelastic modulator (PEM). The reflected signal from the sample is collected through the same objective. It is noteworthy that PL from the heterostructure is negligible compared with the reflected intensity and therefore does not contaminate the DC normalization.

The PEM modulates the polarization state of the reflected light at a frequency of 50~kHz with a maximum retardation of 2.405~rad, enabling sensitive detection of Kerr rotation through phase-sensitive lock-in techniques. The reflected intensity is detected using a silicon avalanche photodiode. The DC component of the signal is recorded using a Keithley source meter, while the AC component is measured using a lock-in amplifier referenced to the second harmonic of the PEM modulation frequency. The Kerr rotation signal is obtained from the ratio $V_{\mathrm{AC}}/V_{\mathrm{DC}}$, where $V_{\mathrm{AC}}$ is the amplitude of the modulated reflection and $V_{\mathrm{DC}}$ the reflected intensity, which is proportional to the Kerr rotation angle. Detection at the second harmonic suppresses background contributions and significantly improves the signal-to-noise ratio, enabling the measurement of small Kerr rotations arising from proximity-induced magnetic interactions.

Spectrally resolved Kerr measurements are obtained by tuning the excitation wavelength across the excitonic resonances of the heterostructure. To isolate the excitonic response of the $\MoSe$ layer, differential MOKE spectra are obtained by subtracting the Kerr spectrum of a bare $\FGT$ reference region from that of the $\MoSe/\FGT$ heterostructure at each excitation energy, which removes the intrinsic Kerr background of the ferromagnetic layer.

\subsection{Fitting of the Kerr spectra}

The differential Kerr spectra are fitted with Eq.~(\ref{eq:lorentzian}) using a least-squares optimization with bounded parameters. Each resonance contributes an amplitude, a center energy and a full width at half maximum, and a single constant offset is shared by the model. Only the center energies and linewidths are bounded. $X_A^{1s}$ is constrained to 1.640--1.700~eV and $X_A^{2s}$ to 1.720--1.790~eV, brackets set by the PL resonance and by the resolved feature at 1.755~eV respectively; $X_B^{1s}$ is constrained to 1.860--1.930~eV and $X_B^{2s}$ to 1.990--2.090~eV, both from the resonances identified in the PLE spectrum. Linewidths are bounded to 10--100~meV, except for $X_B^{2s}$, which lies closest to the renormalized quasiparticle gap and is allowed up to 400~meV. The amplitudes, including their signs, are unbounded, and the optimization is started from a sign-neutral guess so that the initial values cannot impose the polarity that the fit is used to test. The two field polarities are fitted independently. The number of components is chosen by the Akaike and Bayesian information criteria evaluated over models with three to seven resonances; both favor the four-component model reported here. Parameter uncertainties are taken from the diagonal of the covariance matrix returned by the fit. These assume independent, normally distributed residuals and should be read as lower bounds: the point-to-point scatter of the differential spectra exceeds the calculated detection-noise floor by roughly a factor of five, so the residuals are dominated by systematic rather than random contributions.

\subsection{Computation of the absorption spectra}

To understand the experimentally assigned excitonic states, we perform first-principles GW and GW-BSE calculations~\cite{Hybertsen1986,Rohlfing2000}, using the BerkeleyGW package~\cite{Deslippe2012}. Density functional theory calculations are performed using the Quantum ESPRESSO software package~\cite{Giannozzi2009}. In our calculations, we compute the optical absorption of monolayer $\MoSe$. Screening by the metallic $\FGT$ layer is included via a wavevector-dependent correction to the intrinsic dielectric function of $\MoSe$ following previous work~\cite{Chen2015}, parameterized by the inverse screening length $k_0$ (see Supplementary Section~III for details). The A- or B-series character of each calculated exciton peak is determined by the Bloch band projections of the exciton wavefunctions.

\section*{Data availability}

The data that support the findings of this study are available within the paper and its Supplementary Information. The source data underlying the figures are available from the corresponding authors on request.

\section*{Code availability}

The analysis codes used in this study are available from the corresponding authors on request.

\begin{acknowledgments}
This work was supported by JSPS KAKENHI (Grant Nos. JP25K24588, JP25K22210, and JP26K22704), the Canon Foundation, and the Shorai Foundation for Science and Technology. This work was also supported by the Materials Forming Unit of the NIMS Open Facility (NOF) and ARIM (NIMS Nanofabrication). B.P. acknowledges financial support from the Anusandhan National Research Foundation (ANRF) through Grant No. ANRF/ARG/2025/011960/ENS. Z.L. acknowledges the support from United States DOE, Office of Basic Energy Sciences, under the Contract No. DE-SC0026336. An award of computer time was provided by the U.S. Department of Energy's (DOE) Innovative and Novel Computational Impact on Theory and Experiment (INCITE) Program. This research used supporting resources at the Argonne and the Oak Ridge Leadership Computing Facilities. The Argonne Leadership Computing Facility at Argonne National Laboratory is supported by the Office of Science of the U.S. DOE under Contract No. DE-AC02-06CH11357. The Oak Ridge Leadership Computing Facility at the Oak Ridge National Laboratory is supported by the Office of Science of the U.S. DOE under Contract No. DE-AC05-00OR22725. S.P.D. acknowledges the Swedish Research Council (VR) grant (No. 2025-03702). K.W. acknowledges support from the CREST (Grant No. JPMJCR24A5), JST. We thank T. Terashima for assistance with the sample fabrication.
\end{acknowledgments}

\section*{Author contributions}

A.K. fabricated the heterostructures and performed the optical microscopy, photoluminescence, photoluminescence excitation, and magneto-optical Kerr spectroscopy measurements. M.J. performed the reflection measurements and assisted with the optical measurements. B.Z. and Z.L. performed the GW-BSE calculations and the screening-dependent excitonic analyses. D.K. conceived and supervised the project, developed the analysis code, built the measurement systems, and prepared the figures. B.P. supervised the experimental work and co-led the study. Z.W. assisted with the magneto-optical Kerr measurement. K.W. and T.T. grew the hexagonal boron nitride crystals. R.K. contributed measurement resources, and discussions. A.K., B.Z., S.P.D., Z.L., R.K., B.P. and D.K. wrote the manuscript with input from all authors.

\section*{Competing interests}

The authors declare no competing interests.

\bibliography{excitonic_moke}

\end{document}